\documentclass[aps,pre,preprint,letterpaper,superscriptaddress,amsmath,showpacs,floatfix,longbibliography]{revtex4-2}

\usepackage[utf8]{inputenc} 
\usepackage[T1]{fontenc}    
\usepackage{hyperref}       
\usepackage{url}            
\usepackage{booktabs}       
\usepackage{amsfonts}       
\usepackage{nicefrac}       
\usepackage{microtype}      
\usepackage{xcolor}         

\usepackage{enumitem, multirow}
\usepackage{graphicx}
\usepackage{amsmath}
\usepackage{amsthm}
\usepackage{subcaption}
\usepackage{siunitx}
\usepackage{tablefootnote}
\usepackage{tikz}

\begin{document}

\title{Discovery of Symbolic Hamiltonian Expressions with Buckingham-Symplectic Networks}

\date{\today}

\author{Joe Germany}
\affiliation{Department of Mathematics, 
American University of Beirut,
Lebanon}

\author{Joseph Bakarji}
\affiliation{Department of Mechanical Engineering, \\
Complexity and Network Science Cluster, CAMS, \\
American University of Beirut, Lebanon}

\author{Sara Najem}
\affiliation{Department of Physics, \\
Complexity and Network Science Cluster, CAMS, \\
American University of Beirut,
Lebanon}

\begin{abstract}\noindent
Hamiltonian systems lie at the heart of modeling the physical world. Their defining scalar, the Hamiltonian, encodes both energy conservation and symplectic geometry in its phase-space trajectories. Recent deep learning approaches model Hamiltonian systems by embedding their properties either in the architecture or in the loss function. However, they typically ignore that: i) a Hamiltonian carries units of energy and/or ii) that every integrable Hamiltonian admits a canonical transformation to action-angle coordinates in which the dynamics reduce to a simple rotation on an invariant torus. We propose BuSyNet, a deep learning architecture that combines these two constraints via a dimensionally-consistent, symplectic transformation. 
A symplectic layer maps input trajectories to lower-dimensional latent action-angle variables, which are then combined with system parameters to discover a symbolic Hamiltonian expression in units of energy. Evaluated on the harmonic oscillator and the Kepler two‑body problem (in 2D and 3D), BuSyNet recovers concise, closed‑form Hamiltonians that outperform state-of-the-art neural architectures in long‑term prediction accuracy and stability, while maintaining interpretability.

\end{abstract}

\maketitle

\section{Introduction}
Time–series data now permeate every scientific and industrial domain \citep{Faghmous2014BigDataClimate}, from satellite telemetry \citep{Krishnamurthy2015RemoteSensingTS} to single‑cell recordings \citep{Liu2020SingleCellTS} and seismology \citep{yermakov2025seismic}.  
Data-driven approaches for distilling this data into predictive models range from classical spectral methods and interpretable system identification techniques to foundation models that are fueled by recent successes in language and vision, leveraging computational parallelization and scaling laws.
Yet many machine learning techniques still remain physics‑agnostic, ignoring conservation laws, symmetries, and physical units that strongly constrain data generated by physical processes and are essential for robust generalization. Recent calls for standardized evaluation protocols in scientific machine learning, such as the Common Task Framework (CTF) \citep{wyder2025common}, further emphasize the need for principled and comparable assessment of physically structured models.



Physics-informed machine learning (PIML) incorporates prior knowledge either as soft constraints in the loss (e.g.\ physics-informed neural networks (PINNs) minimizing differential equation residuals \citep{raissi2019physics}) or as hard architectural constraints, such as symplectic networks for Hamiltonian systems \citep{jin2020sympnets}, equivariant neural networks enforcing symmetry groups \citep{cohen2016group, cohen2019gauge, kondor2018clebsch, decelle2019learning}, and unit-aware architectures such as BuckiNet that embed dimensional consistency directly into the network structure \citep{Bakarji2022}. Complementary approaches attempt to discover governing equations directly from data. Notable examples include Sparse Identification of Nonlinear Dynamics (SINDy) and its variants \citep{brunton2016discovering, rudy2017data, fasel2022ensemble, messenger2021weak}, genetic algorithms for symbolic regression \citep{schmidt2009distilling}, and Kolmogorov-Arnold Networks \citep{liu2024kan}. 

Beyond explicit equation discovery, a growing body of work focuses on uncovering invariants and latent struture from trajectories. Neural approaches have been used to learn independent conservation laws \citep{zhu2023machine, liu2022machine, hou2026machine, fang2025constants}, invariants from observed dynamics \citep{liu2021machine}, and hidden symmetries in dynamical systems \citep{liu2022machine, decelle2019learning}, demonstrating that learning invariants can provide a more compact representation of the physical dynamics than modeling full vector fields.


Modeling physical time-series data remains challenging due to high dimensionality, nonlinearity, noise, and hidden variables. Several recent approaches therefore seek coordinate transformations or embeddings in which the dynamics simplify. 
Koopman-inspired methods lift nonlinear flows to higher-dimensional linear representations \citep{lusch2018deep}. This viewpoint has motivated extensions of Dynamic Mode Decomposition (DMD), such as extended DMD \citep{williams2015data} and Koopman autoencoders, although the precise relationship between DMD and the Koopman operator remains nuanced and depends on the choice of observables and approximation space \citep{tu2014dynamic, balabane2021koopman}.
Other methods learn low-dimensional latent manifolds constrained by polynomial differential equations, including SINDy autoencoders \citep{champion2019data}, deep delay autoencoders \citep{bakarji2023discovering}, and latent manifold dynamics models \citep{fries2022lasdi}. In integrable Hamiltonian systems, such transformations can be taken further: structure-preserving transformation of the position and momentum coordinates to action–angle variables, in which motion reduces to uniform rotation on invariant tori. In these coordinates, the actions remain constant while the angles evolve linearly in time. More generally, for conservative systems, the Hamiltonian provides a fundamental invariant, a scalar quantity governing both energy conservation and the symplectic geometry of phase-space trajectories.



Hamiltonian neural networks (HNNs) \citep{greydanus2019hamiltonian} explicitly exploit geometric structure by learning a Hamiltonian function whose gradients define the dynamics. Beyond modeling physical systems, Hamiltonian-based energy-preserving networks have also been applied to image prediction \citep{greydanus2019hamiltonian}, generative modeling \citep{toth2019hamiltonian}, and optimal control \citep{meng2022sympocnet, zhang2024time, zhong2019symplectic}, demonstrating the broader utility of structure-preserving dynamics in machine learning. Subsequent work has enforced symplectic structure directly in network architectures \citep{jin2020sympnets, tapley2024symplectic, burby2020fast}, incorporated symplectic integrators into training \citep{chen2019symplectic, zhong2019symplectic, sanchez2019hamiltonian, tong2021symplectic, xiong2020nonseparable}, and extended the framework to structured settings such as adaptable Hamiltonian networks \citep{han2021adaptable}, port-Hamiltonian neural networks for time-dependent systems \citep{desai2021port}, separable Hamiltonian neural networks \citep{khoo2024separable}, and constrained or augmented formulations incorporating Dirac constraints and external interactions \citep{kaltsas2025constrained, li2025augmented}. Recent studies have further analyzed symplectic structure preservation and numerical stability properties of learned Hamiltonians \citep{david2023symplectic, horn2025generalized}, as well as applications to domain-specific physical problems \citep{cipriani2025hamiltonian}. Approaches involving action-angle coordinates have also been explored in \citep{daigavane2022learning, ishikawa2021neural}, allowing networks to learn transformations that simplify dynamics and reveal underlying integrable structure. These works consistently demonstrate that embedding geometric structure improves long-term stability and interpretability. 

Despite these advances, comparatively little attention has been paid to dimensional consistency within Hamiltonian learning. The Hamiltonian is not merely an invariant scalar; it must carry units of energy and be expressible in terms of dimensionally consistent combinations of physical quantities. Existing Hamiltonian neural networks typically mix variables of heterogeneous physical dimensions without enforcing consistency, potentially impairing interpretability and extrapolation. Recent work on unit-aware learning architectures demonstrates that embedding dimensional analysis directly into network structure improves generalization and interpretability \citep{Bakarji2022}. However, such approaches have not yet been systematically integrated with symplectic and Hamiltonian learning frameworks.


We close this gap with Buckingham-Symplectic Networks (BuSyNet), a two‑stage architecture that marries symplectic geometry with dimensional consistency to map time-varying measurements to a constant of motion with physical units.

A learnable symplectic transformation, inspired by SympNet \citep{chen2019symplectic}, first maps the observed $(q,p)$ trajectories to a latent action–angle space, ensuring canonical structure and reducing the dynamics to linear motion on an invariant torus.  
A subsequent unit‑aware layer, inspired by the Buckingham‑$\pi$ theorem and building on BuckiNet \citep{Bakarji2022}, combines these latent variables with known system parameters to yield a symbolic expression for the Hamiltonian that is dimensionally consistent by construction. 
The model therefore transforms a raw, time‑dependent sequence into action-angle coordinates with constant actions, and subsequently into a single time‑invariant constant of motion expressed in physically meaningful units. 

We demonstrate that this synergy of symplectic structure and unit‑awareness delivers interpretable Hamiltonians with superior long‑term prediction accuracy on canonical benchmarks (harmonic oscillator and Kepler orbits), outperforming HNNs, SympNets, and vanilla neural network baselines.
BuSyNet reduces long‑term rollout error on both benchmarks while returning closed‑form Hamiltonians that match the analytic ground truth.


\section{Theoretical Background}
Consider measurement time series assumed to be generated by an integrable Hamiltonian system with states $z(t) = (\mathbf p(t), \mathbf q(t)) \in \mathbb R^{2n}$ and an associated Hamiltonian $H(\mathbf q, \mathbf p, t)$, which satisfies Hamilton's canonical equations
\begin{equation}\label{eq:ham}
    \dot{\mathbf{q}} =
    \dfrac{d \mathbf{q}}{dt} =
    \dfrac{\partial H(\mathbf{q}, \mathbf{p})}{\partial \mathbf{p}}, \quad
    \dot{\mathbf{p}} = \dfrac{d\mathbf{p}}{dt} = - \dfrac{\partial H(\mathbf{q}, \mathbf{p})}{\partial \mathbf{q}}
\end{equation}
or, more succinctly, 
\begin{equation}\label{eq:shortham}
    \dot{\mathbf{z}} = \mathbf{J} \nabla_{\mathbf{z}} H(\mathbf{z}), \quad \text{s.t.} \quad \mathbf{J} = \begin{bmatrix} 0 & I_n \\ - I_n & 0 \end{bmatrix},
\end{equation}
where $\mathbf J$ is called the symplectic matrix (with $\mathbf{J}^{-1} = \mathbf{J}^\top = - \mathbf{J}$), and $I_n \in \mathbb R^{n\times n}$ is the identity matrix. 
As a consequence, a differentiable map $\Phi: U \to \mathbb{R}^{2n}$ is \emph{symplectic} if it satisfies the condition
\begin{equation}
        \left(\dfrac{\partial \Phi}{\partial \mathbf{z}} \right)^\top \mathbf{J} \left(\dfrac{\partial \Phi}{\partial \mathbf{z}} \right) = \mathbf{J},
\end{equation}
    where $\partial \Phi/\partial \mathbf{z}$ is the Jacobian matrix of $\Phi$.
If $\phi_t (\mathbf{z}_0) = \mathbf{z} (t)$ is the \emph{phase flow} of a \emph{Hamiltonian system}, which is the solution to the system of $2n$ differential equations $\dot{\mathbf{z}} = \mathbf{J} \nabla_\mathbf{z} H(\mathbf{z})$ with the initial condition $\mathbf{z}_0 = (\mathbf{q}_0, \mathbf{p}_0)$, then the \emph{phase flow} $\phi_t$ is \emph{symplectic}. (see Theorem 2.4 in \citep{hairer2013geometric} (p. 184-185))

This formalism entails that there exists a (generally unknown) \textbf{symplectic map} that sends $(\mathbf q, \mathbf p)$ into action-angle coordinates ($\mathbf I, \boldsymbol \theta)$, where $\mathbf I$ is the action (with units of $ML^2/T$, i.e. Energy$\times$Time) and $\boldsymbol \theta$ is the angle. In these coordinates, the actions $\mathbf I(t)$ remain constant in time, while the angles $\boldsymbol \theta(t)$ evolve linearly, such that 
\begin{equation}
    \boldsymbol \theta(t + \Delta t) = \boldsymbol \theta (t) + \boldsymbol{\dot \theta} \Delta t,
\end{equation}
where $\boldsymbol{\dot \theta}= \partial H(\mathbf I)/\partial \mathbf I$.
Furthermore, the Hamiltonian has \textbf{units of energy} (i.e. Joules, or in terms of base units, $ML^2/T^2$, where $M$ is Mass, $L$ is length, and $T$ is time). Consequently, the functional expression of $H$ must be dimensionally consistent with the physical parameters of the system, such as mass, spring constants, or gravitational constant.

Given data pairs for the position ($\mathbf{q}$) and momentum ($\mathbf{p}$) of an integrable system $\{(\mathbf q_i, \mathbf p_i)\}_{i=1}^N$, physically measured parameters that characterize the system, which we denote by $\{\mathbf m_i\}_{i=1}^M$ (such as the mass and spring constant for harmonic oscillators), and dimension matrices characterizing the units of the variables and parameters $\{D_{\text{in}}, D_{\text{out}}\}$, our learning task is three-fold:

\begin{enumerate}[label=\textbf{Aim \arabic*}, left=0pt]

    \item \phantomsection\label{obj:discover_itheta}
    Learn the map $(\mathbf{q}_k, \mathbf{p}_k) \mapsto (\mathbf I_k, \theta_k)$ that encoders the state into a action-angle space with a constant $I$ and a time-dependent $\theta$ , where the dynamics evolve on an invariant torus. 

    \item \phantomsection\label{obj:evolve}
    Learn the nonlinear transformation $(\mathbf{q}_k, \mathbf{p}_k) \mapsto (\mathbf{q}_{k+1}, \mathbf{p}_{k+1})$ to evolve the systems states one time step forward. This map enables the method to be used recursively for temporal inference.

    \item \phantomsection\label{obj:discover}
    Discover a symbolic, dimensionally consistent expression for the Hamiltonian quantity of the system as a function of the actions $\mathbf I$ and the system parameters: $\hat H(\mathbf I, \mathbf m)$.
\end{enumerate}

\begin{figure}[h!]
  \centering
  \includegraphics[width=0.8\linewidth]{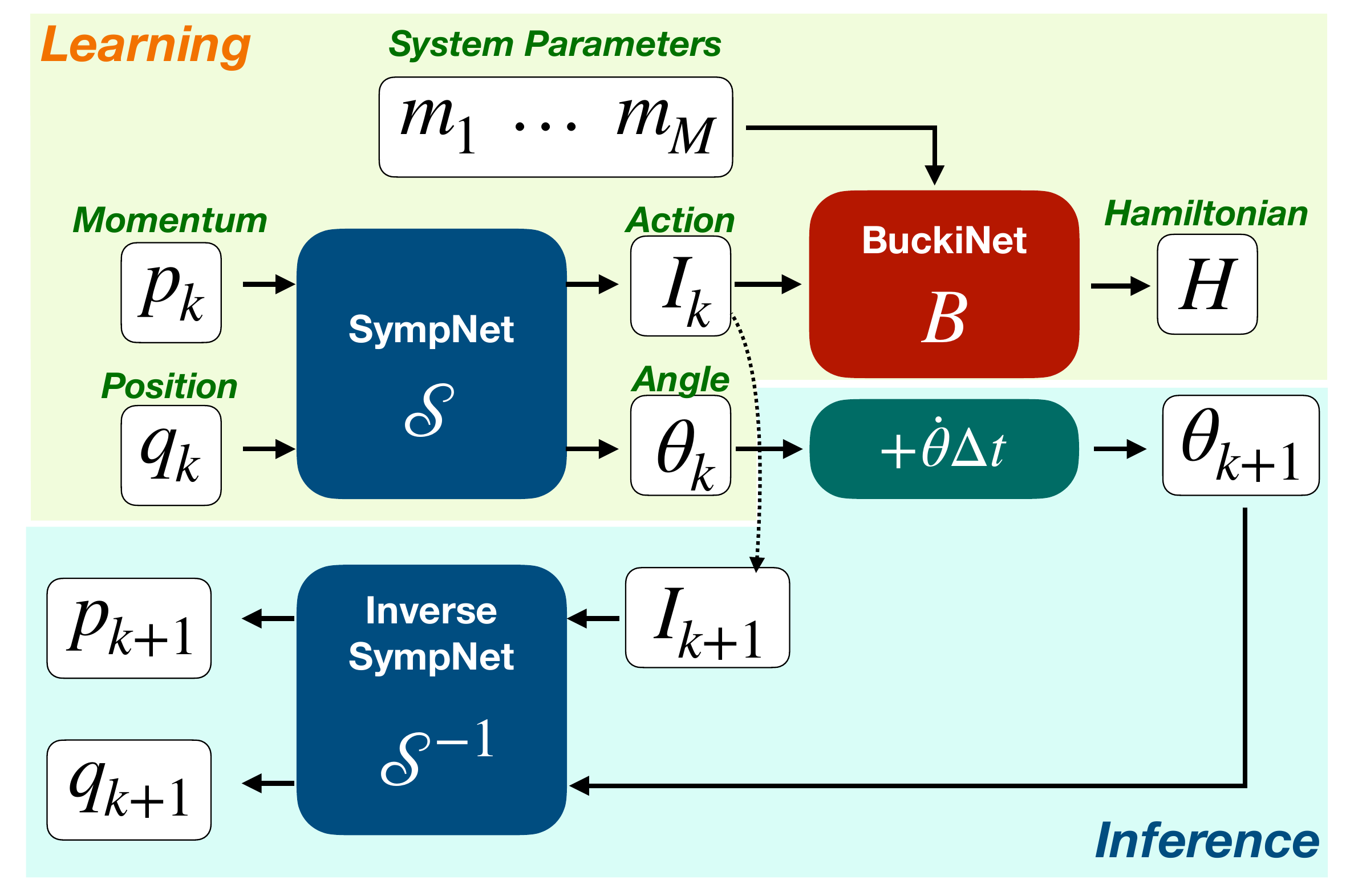}
  \caption{Illustration of the BuSyNet architecture. Measurements of position ($q_k$) and momentum ($p_k$) at time index $k$ are fed into a symplectic network, transformed into actions $I_k$ and angles $\theta_k$. Actions are then fed into a BuckiNet head to reconstruct an analytical expression for the Hamitonian $H$ with known equations of motion. At inference, an Euler step is applied to  $\theta_k$, while $I_k$ stays the same, and $p_{k+1}$ and $q_{k+1}$ are obtained via the inverse of SympNet.}
  \label{fig:architectureDiagram1}
\end{figure}

These objectives are achieved through a combination of a physics-informed loss (inspired by physics-informed neural networks \citep{raissi2019physics}), and a physics-informed architecture (building on BuckiNet \citep{Bakarji2022} and SympNet \citep{jin2020sympnets}). Figure~\ref{fig:architectureDiagram1} shows an illustration of the method, designed to accomplish the above aims and examined more closely in the below section.


\section{Buckingham-Symplectic Networks (BuSyNet)}

\subsection{Background}
BuSyNet is built from two prior ideas: \emph{symplectic neural maps} that keep Hamiltonian structure intact, and \emph{unit–aware layers} that enforce dimensional consistency.  We summarize both in turn and highlight the extensions required for Hamiltonian discovery.

\paragraph{Symplectic maps via (G-)SympNet.}
SympNets \citep{jin2020sympnets} mimic the leap–frog updates of classical splitting integrators: each layer alternately shifts $\mathbf q$ by a function of $\mathbf p$ and $\mathbf p$ by a function of $\mathbf q$. 
Because every layer is symplectic, their composition is a canonical transformation by construction.  
Earlier work used SympNets to learn time-stepping flows, i.e.\ $(q(t),p(t))\!\mapsto\!(q(t+\Delta t),p(t+\Delta t))$ to be used recursively for temporal prediction \citep{jin2020sympnets}.
Our goal is different: we train a SympNet encoder $\mathcal S$ that pushes raw phase-space coordinates into \emph{action–angle} variables \((\mathbf I,\theta)\), which also happens to be a symplectic map, as done in \citep{daigavane2022learning}.
In that latent space actions are constant and angles advance uniformly, turning $2n$ dimensional complex trajectories into straight lines on an $n$-torus.  
This not only stabilizes long-horizon rollouts but also yields a separable Hamiltonian \(H(\mathbf I)\) that is easier to recover symbolically.

\paragraph{Dimensional consistency via BuckiNet.}

BuckiNet layers provide a neural implementation of the Buckingham-\(\pi\) theorem~\citep{Bakarji2022}.
They transform input measurements of system variables/parameters with known units (e.g. Kg, Newton, Joules, etc.) to dimensionless (i.e. unit-less) latent variables as part of a deep network architecture. 
For example, given measurements of density, viscosity, velocity and length, in the context of a fluid flow simulation, they might discover the Reynolds number, which can then be used by subsequent layers to predict system behavior while ensuring that the learned representations respect the underlying physical scaling laws. This allows the network to generalize across different regimes and parameter scales, improving interpretability and robustness. 

Let each input variable $v_i$ carry a units (dimension) vector  $\Omega(v_i)\in\mathbb Z^{r}$ over the $r$ base dimensions (e.g.\ $M,L,T$).
BuckiNet encodes a transformation of the form $\pi_i = \prod v_j^{{\Psi}_{ij}}$, where $\Psi_{ij}$ are the fitting parameters of the layer, and $\pi_i$ is the $i-th$ output of the layer.
Stacking the the units vectors gives the \emph{dimensional matrix} $D_{\text{in}}\!=\![\Omega(v_1)\;\cdots\;\Omega(v_d)]\in\mathbb Z^{r\times d}$. 
BuckiNet converts the inputs to \emph{log space}, $\tilde{\boldsymbol V}=\log|\boldsymbol V|\in\mathbb R^{m\times d}$, applies a \emph{linear} map with trainable exponents $\Psi\in\mathbb R^{d\times d'}$,  and exponentiates back:

\begin{equation}
    \boldsymbol{\Pi} \;=\; 
\exp\!\bigl(\,\tilde{\boldsymbol V}\,\Psi\,\bigr)
\;=\;
\left[\,
\prod_{i=1}^{d}v_i^{\Psi_{i1}},\;
\ldots,\;
\prod_{i=1}^{d}v_i^{\Psi_{id'}}\right].
\end{equation}

While BuckiNet has been previously used to encode dimensional inputs into dimensionless latent variables, we propose an extension to provide outputs \(\pi_i\) with specific dimensions, according to the following constraint
\begin{equation}\label{eq:buckinet_const}
    D_{\text{in}}\Psi \;=\; D_{\text{out}},
\end{equation}
where \(D_{\text{out}}\) encodes the desired units (\(D_{\text{out}}=\mathbf 0\) recovers the classical dimensionless null-space). 
In practice, Eq. (\ref{eq:buckinet_const}) is enforced with a soft penalty  
$\lambda\|D_{\text{in}}\Psi-D_{\text{out}}\|^2_2$, so the exponents $\Psi$ are learned jointly with the rest of the network through back-propagation. 
After training, each column of $\Psi$ provides an explicit monomial; hence, in our setting, a \emph{symbolic Hamiltonian} in correct energy units.

In this problem, the input variables of interest are the actions $\mathbf I$ and the system parameters $\mathbf m$. The output a symbolic expression for the Hamiltonian combining elements in $\mathbf I$ and $\mathbf m$.

\subsection{Architecture}
\paragraph{Grounding Action Calculations.}
To enable SympNet to learn the unique transformation mapping $(\mathbf q, \mathbf p)$ to action-angle coordinate $(\mathbf I, \boldsymbol \theta)$ that preserve the phase-space dynamics, we first compute the numerical values of the actions $\mathbf I$. Each action $I_i$ is defined via the classical integral
\begin{equation}
\label{actions}
    I_i = \frac{1}{2\pi} \oint p_i \ dq_i = \frac{1}{2\pi} \int_0^{T_i} p_i \frac{dq_i}{dt} \ dt,
\end{equation}
where $T_i$ is the period of the motion for the $i$-th degree of freedom. The period $T$ is estimated by applying a Fast Fourier transform to the original trajectory data, and the integral is evaluated numerically using Simpson's quadrature rule. This step is essential, otherwise there are too many unknowns for the loss function to discover (given the functional form we adopt below for the Hamiltonian) and we will not discover the true underlying action-angle transformation.

\paragraph{Unit-Aware BuckiNet Head.}
Next, we train the BuckiNet network to discover the correct exponents of the actions and the input parameters, prior to learning the symplectic transformation. The precomputed actions $\mathbf I$ and physical parameters $\mathbf m$ are fed into a \textit{BuckiNet} layer $B$, which learns the mapping:
\begin{equation}
\label{latent_functional_form}
    \hat H
     =B(\mathbf I,\mathbf m)
     = \Bigl[\sum_{j} I_j\Bigr]^\alpha \prod_{i} m_i^{\beta_i},
\end{equation}
where the exponents are $\Psi=[\alpha, {\boldsymbol \beta}]$. The BuckiNet head is trained to satisfy $D_{\text{in}}\Psi=D_{\text{out}}$ (i.e. \ref{obj:discover}), with $D_\text{out}$ encoding energy units, via the soft constraint:
\begin{equation}
    \mathcal L_\text{dim} =  \bigl\| D_{\text{in}}\Psi-D_{\text{out}} \bigr\|_F^2.
\end{equation}
Eq.~\ref{latent_functional_form} serves as a latent functional form: it is not the final Hamiltonian but it is used to extract the physically-consistent exponent $\alpha$ for the actions. Once $\alpha$ is determined, we define the Hamiltonian in a more flexible form for SympNet training that includes all possible monomials of the actions $\mathbf I$ with total degree $\alpha$:
\begin{equation}
\label{full_action_combination_H}
\begin{cases}
    \hat H (\mathbf I) = \sum_{\mathbf{k} \in \mathcal{K}_\alpha} a_{\mathbf{k}} \, I_1^{k_1} I_2^{k_2} \cdots I_n^{k_n}, &\alpha>0,\\[8pt]
    \dfrac{1}{{\hat H (\mathbf I)}} = \sum_{\mathbf{k} \in \mathcal{K}_{\alpha}} a_{\mathbf{k}} \, I_1^{k_1} I_2^{k_2} \cdots I_n^{k_n}, &\alpha<0,    
\end{cases}
\end{equation}
where $\mathbf{k} = (k_1, \dots, k_n)$ is a multi-index with $\sum_{j=1}^n k_j = |\alpha|$, $\mathcal{K}_\alpha$ is the set of all such multi-indices, and $\{a_{\mathbf{k}}\}$ are learnable coefficients corresponding to each allowed monomial. 
These coefficients are optimized during subsequent SympNet training, allowing the network to capture all unit-consistent combinations of the actions.
This formulation removes the assumption of summing the actions as in Eq.~\ref{latent_functional_form} and allows for additional unit-free prefactors that could not be determined in the latent form because they require access to the trajectory data for learning.

\paragraph{SympNet Losses.}
To satisfy \ref{obj:discover_itheta}, we learn a strictly symplectic, invertible map
$\mathcal S_\phi\!:\;(\mathbf q,\mathbf p)\mapsto(\mathbf I,\boldsymbol\theta)$ using the (G-)SympNet architecture \citep{jin2020sympnets}. $\phi$ are the learning parameters of the SympNet encoder. The architecture here consists of a SympNet network, which then feeds in the learned actions $\hat {\mathbf I}$ into the unit-correct Hamiltonian functional form \eqref{full_action_combination_H}.

Since each layer of a SympNet is canonical, $\mathcal S$ guarantees that $\hat{\mathbf I}$ are approximate constants of motion while $\hat{\boldsymbol\theta}$ advance almost linearly, providing a latent torus on which long-horizon prediction is trivial. Constraints specific to the equations of motion can be inferred from the architecture. In particular, the equalities $\partial H/\partial \mathbf I = \dot{\boldsymbol \theta}$, and $\partial H/\partial \boldsymbol \theta = - \dot{\boldsymbol I}$ have to hold, as well as the equality of the predicted action output from SympNet and the calculated action values from Step 1 above, entailing to the following losses:
\begin{equation}
    \mathcal L_{\mathbf{I}} = \left\lVert \mathbf{I} - \hat{\mathbf{I}} \right\rVert _2 ^2, \qquad 
    \mathcal L_{{\dot{\boldsymbol \theta}}} = \left\lVert \dfrac{\partial \hat{H}}{\partial \hat{\mathbf{I}}} - \dfrac{d \hat{\boldsymbol{\theta}}}{dt} \right\rVert _2 ^2, \qquad 
    \mathcal L_{{\dot{\boldsymbol I}}} = \left\lVert \dfrac{\partial \hat{H}}{\partial \hat{\boldsymbol{\theta}}} + \dfrac{d \hat{\mathbf{I}}}{dt} \right\rVert _2 ^2,
\end{equation}
where $\partial \hat{H} (\hat{\mathbf{I}})/\partial \hat{\mathbf{I}}$ and $\partial \hat{H} (\hat{\mathbf{I}})/\partial \hat{\boldsymbol{\theta}}$ are computed via auto-differentiation \citep{baydin2018automaticdifferentiationmachinelearning}. Furthermore, 
\begin{equation}
\dfrac{d \hat{\mathbf{I}}}{dt} = \dfrac{\partial \hat{\mathbf{I}}}{\partial \mathbf{q}} \dfrac{d \mathbf{q}}{dt} + \dfrac{\partial \mathbf{\hat{\mathbf{I}}}}{\partial \mathbf{p}} \dfrac{d \mathbf{p}}{dt}, \quad
\text{and}  \quad \dfrac{d \hat{\boldsymbol{\theta}}}{dt} = \dfrac{\partial \hat{\boldsymbol{\theta}}}{\partial \mathbf{q}} \dfrac{d \mathbf{q}}{dt} + \dfrac{\partial \mathbf{\hat{\boldsymbol{\theta}}}}{\partial \mathbf{p}} \dfrac{d \mathbf{p}}{dt},
\end{equation}
where $\partial \hat{\mathbf{I}}/\partial \mathbf q$, $\partial \hat{\mathbf{I}}/\partial \mathbf p$, $\partial \hat{\boldsymbol \theta}/\partial \mathbf q$, and $\partial \hat{\boldsymbol \theta}/\partial \mathbf p$ are also all computed via automatic differentiation, while $\dot {\mathbf p}$ and $\dot {\mathbf q}$ are obtained by numerically differentiating the input. $d\hat{\boldsymbol\theta}/dt$ is automatically zero (by construction), since $\hat {\boldsymbol\theta}$ is not present in the Hamiltonian functional form \eqref{full_action_combination_H}.




\paragraph{Training loss.}
We jointly optimize the SympNet weights, $\phi$, and $\hat H$ functional form coefficients, $\{a_\mathbf{k}\}$, with the objective
\begin{align}
\mathcal L(\phi, \Psi)
    &= \lambda_{\mathbf{I}} \mathcal L_{\mathbf I}
    +\lambda_{{\dot{\boldsymbol \theta}}} \mathcal L_{{\dot{\boldsymbol \theta}}}
    + \lambda_{\mathbf{\dot{I}}} \mathcal L_{\mathbf{\dot{ I}}}.
\label{eq:loss}
\end{align}
The loss hyperparameter weights are all chosen to be $1$ in this study, such that $\lambda_{\mathbf{I}} = \lambda_{{\dot{\boldsymbol \theta}}} = \lambda_{{\dot{\boldsymbol I}}} = 1$.

\paragraph{Latent-space rollout (forecasting).}
Once trained, $\dot{\boldsymbol{\theta}}$ is obtained by auto-differentiating the Hamiltonian output $\hat H (\mathbf{I}, \mathbf m)$ of the training data with respect to the learned actions $\mathbf{I}$; given that $\dot{\boldsymbol{\theta}}$ is constant in time. This procedure is shown in Figure \ref{fig:architectureDiagram1}.
Thus, we achieve \ref{obj:evolve} by recursively perfoming the following latent-space update: 

\[
\begin{array}{c @{\hspace{0.5cm}} c @{\hspace{0.5cm}} c}
(\mathbf q_k,\;\mathbf p_k)
&
&
(\mathbf q_{k+1},\;\mathbf p_{k+1})
\\[8pt]
\Big\downarrow\;{ \mathcal S_{\phi}}
&
&
\Big\uparrow\;{ \mathcal S_{\phi}^{-1}}
\\[1pt]
(\hat{\mathbf I}_k,\;\hat{\boldsymbol\theta}_k)
&
\overset{\displaystyle
      \hat{\boldsymbol\theta}_{k+1} = 
      \hat{\boldsymbol\theta}_k + \frac{d\hat{\boldsymbol\theta}}{dt}\, \Delta t}
      {\underset{\displaystyle
      \hat{\mathbf I}_{k+1} =\, \hat{\mathbf I}_k}{\xrightarrow{\hspace{4cm}}}}
&
(\hat{\mathbf I}_{k+1},\;\hat{\boldsymbol\theta}_{k+1})
\end{array}
\]

\noindent where the gradient $d\hat{\boldsymbol \theta}/dt = \partial \hat H/\partial \mathbf{I}$ (found above) and $\mathcal S_{\phi}^{-1}$ is computed analytically via layer-wise inversion of SympNet blocks, as shown in Figure \ref{fig:architectureDiagram1}.

\paragraph{Architecture Parameters.}
Initially, we compute the action integrals according to equation \eqref{actions}. Then, we run the L-BFGS optimization algorithm (second-order gradient descent algorithm) for 100 epochs for BuckiNet with the dimensional loss term only, so that we get the dimensionally correct exponents for the actions in the expression of the Hamiltonian, rounding them to the nearest thousandth. 

Next, we initialize a G-SympNet architecture with 4 layers and 32 neurons each, together with the Hamiltonian functional form \eqref{full_action_combination_H}. The weights of SympNet and the set of learnable coefficients $\{a_\mathbf{k}\}$ in the Hamiltonian are trained using the Adam optimizer for 1000 epochs with a learning rate of $10^{-3}$ and the loss function \eqref{eq:loss} where the hyperparameters are all set to $1$ ($\lambda_\mathbf{I} = \lambda_{\dot{\boldsymbol\theta}} = \lambda_{\mathbf{\dot I}} = 1$).

\section{Experimental Results}
We benchmark BuSyNet on two classical, analytically integrable Hamiltonians that are standard in the literature: the 1D simple harmonic oscillator and the Kepler two-body problem in 2D and 3D. In each case we compare against a vanilla feed-forward network (NN), Hamiltonian Neural Networks (HNN) \citep{greydanus2019hamiltonian}, and SympNet \citep{jin2020sympnets}.

\subsection{Harmonic Oscillator}
A mass $m$ connected to a spring of stiffness $k$ has Hamiltonian
\begin{equation}\label{eq:shmham}
        H(q,p)=\frac{1}{2}k\,q^{2}+\frac{p^{2}}{2m},
\qquad
    \dot q = \frac{p}{m},\;
    \dot p = -kq .
\end{equation}
For this system the canonical transformation to action–angle variables yields
\begin{equation}\label{eq:shm-ham-aa}
       H(I)=\omega I,
    \qquad
    \omega=\sqrt{k/m}. 
\end{equation}
We generate a ground-truth trajectory with $k=m=1$ and a time step $\Delta t = 0.001$ from $t = 0$ to $t = 20 \pi \ ( = 10 \times \text{period})$. Our training data consists of 1000 evenly spaced points from $t = 0$ to $t = 7$, and we test (for long-term prediction) on the rest.

\subsection{Kepler Problem}
The Kepler problem models a body of mass $m$ orbiting a central mass $M$ under the inverse-square potential $V(r)=-k/r$ with $k=GMm$, where $G$ is the gravitational constant.  
In spherical coordinates \( \mathbf q=(r,\theta,\phi) \) and conjugate momenta \( \mathbf p=(p_r,p_\theta,p_\phi) \), the Hamiltonian is written as
\begin{equation}\label{eq:keplerham}
    H(\mathbf q,\mathbf p)=\frac{1}{2m}\!\left(
    p_r^{2}+\frac{p_\theta^{2}}{r^{2}}
    +\frac{p_\phi^{2}}{r^{2}\sin^{2}\theta}\right)
    -\frac{k}{r}.
\end{equation}
A lengthy analytic action–angle transformation reduces this to
\begin{equation}\label{eq:kepler-ham-aa}
    H(\mathbf I)= -\frac{mk^{2}}
    {2(I_r+I_\theta+I_\phi)^{\,2}} = - \dfrac{mk^2}{2\mathcal{I}^2},
\end{equation}
with $\mathcal{I} = I_r + I_\theta + I_\phi$. We expect 
\begin{equation}\label{eq:freqs}
\omega_i = \frac{\partial H}{\partial I_i} = \frac{d H}{d \mathcal{I}} \cdot \frac{\partial \mathcal I}{\partial I_i} = \frac{d}{d \mathcal I} \left(- \frac{m k^2}{2 \mathcal I^2} \right) = \frac{m k^2}{\mathcal I^3}.
\end{equation}
For Kepler 3D, we generate a ground-truth trajectory with $k=m=1$, a time step $\Delta t = 0.001$ from $t = 0$ to $t = 20$ and initial data $r(0)=0.25, \, \theta(0)=\frac{\pi}{2}, \, \phi(0)=0, \, p_r(0)=0, \, p_\theta(0)=0, \, p_\phi(0)=0.5$ corresponding to a circular trajectory of radius $0.25$ m. Our training data consists of the $10\text{,}000$ evenly spaced points from $t = 0$ to $t = 10$, and we test on the interval from $t=0$ to $t=20$. Note that the period of the system is $T=\frac{\pi}{4}$. We do not assume this for training; we just use it during evaluation to check the extent of generalization.

Similarly, for Kepler 2D, the Hamiltonian expression simplifies as the variables $\phi$ and $p_\phi$ are absent. For our architecture, we use the same parametric values and setup, but with the initial data $r(0)=0.25, \, \theta(0)=0, \, p_r(0)=0, \, p_\theta(0)=0.5$. We also train on $t=0$ to $t=10$ and test on $t=0$ to $t=20$. Similarly, the period can be computed to be $T=\frac{\pi}{4}$.

For all three systems above, we compare BuSyNet against Hamiltonian Neural Network (HNN), SympNet, and Vanilla neural network baselines on long-term rollout error, energy drift, and symbolic-recovery accuracy where applicable.


\subsection{Discovered Hamiltonians}

BuSyNet recovers the following correct, symbolic Hamiltonians
\begin{align}
    \hat H_\text{SHM}(I) &= \dfrac{I^{1.0000} \times k^{0.5000}}{m^{0.5000}} \approx \text{Eq.}~\eqref{eq:shm-ham-aa} \\
    \hat H_\text{Kepler 2D}(\mathbf{I}) &= - \dfrac{1}{2.000 \times (I_r + I_\theta)^{2.000}} \approx \text{Eq.}~\eqref{eq:kepler-ham-aa}, \\
    \hat H_\text{Kepler 3D}(\mathbf{I}) &= - \dfrac{1}{2.000 \times (I_r + I_\theta + I_\phi)^{2.000}} \approx \text{Eq.}~\eqref{eq:kepler-ham-aa},
\end{align}
which closely match Eqs.~(\ref{eq:shm-ham-aa}) and (\ref{eq:kepler-ham-aa}). The results are also compatible with the analytical forms up to numerical rounding. For the simple harmonic oscillator, our recovered equation displays exact dependence on the parameters $k$ and $m$, because we use BuckiNet as the functional form itself, not the form \eqref{full_action_combination_H}. This is in fact usually possible for one-dimensional systems, as their Hamiltonian in action-angle space will be a monomial. However, for the Kepler problem, we use the functional form \eqref{full_action_combination_H}, since in general the Hamiltonian might consist of many terms, to keep our framework general and applicable to higher dimensionality problems.

\subsection{Forecast Accuracy}
Vanilla neural networks were trained to predict $(q(t+\Delta t), p(t + \Delta t))$ given $(q(t),p(t))$. While for Hamiltonian Neural Networks, a symplectic integrator (Leapfrog) is used for inference (\citep{chen2019symplectic}).

\begin{figure}[htbp]
  \centering
  \begin{minipage}{\textwidth}
    \centering
    \begin{subfigure}[t]{0.47\textwidth}
        \centering
        \includegraphics[width=\textwidth]{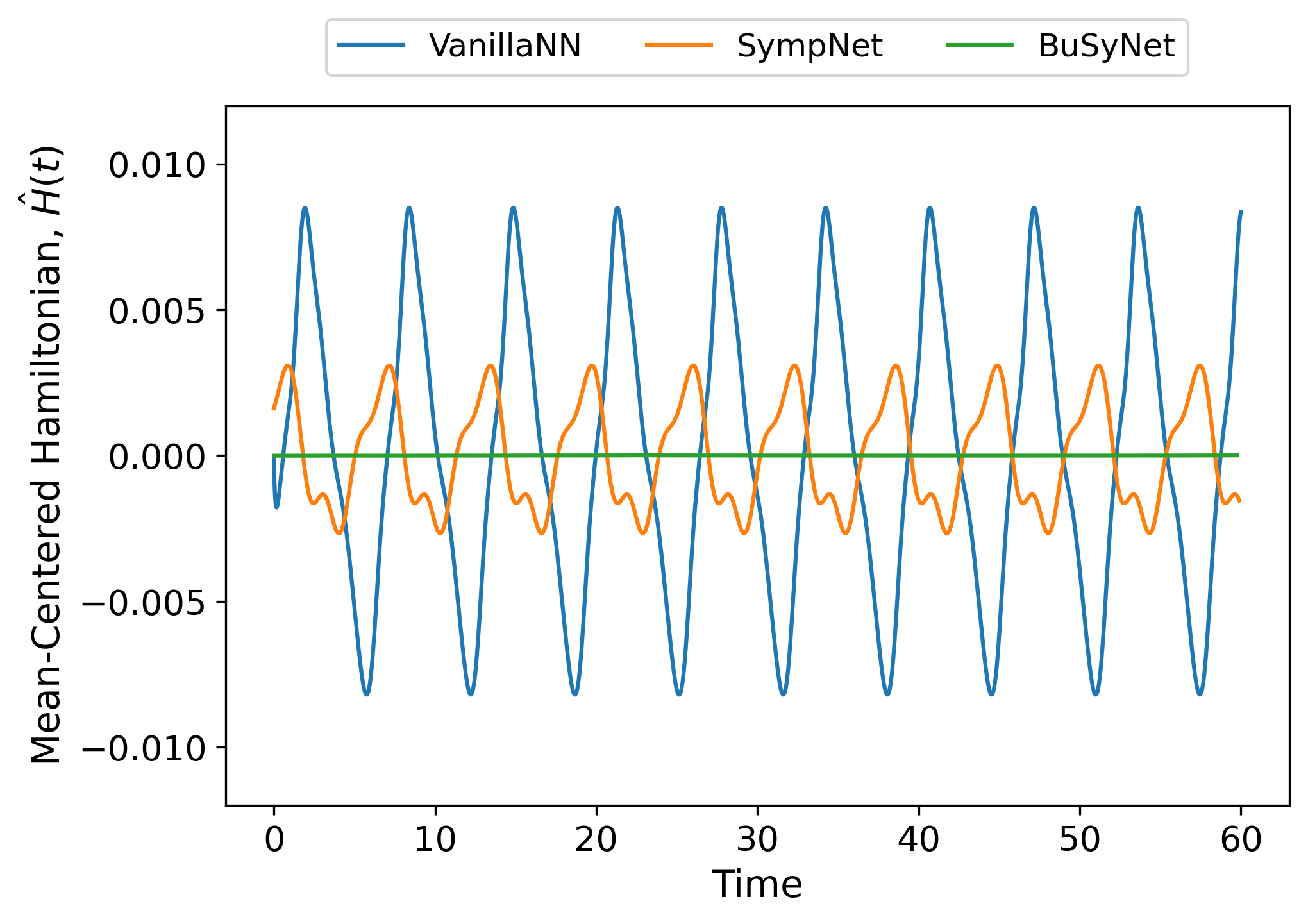}
        \caption{Mean-centered Hamiltonian for the harmonic oscillator across the different methods, showcasing the superior behavior of BuSyNet. HNN is omitted due to its high variance, which exceeds the chosen display scale.}
        \label{fig:SHMHamiltonianPrediction}

    \end{subfigure}
    \hspace{0.05\textwidth}
    \begin{subfigure}[t]{0.45\textwidth}
        \centering
        \includegraphics[width=\textwidth]{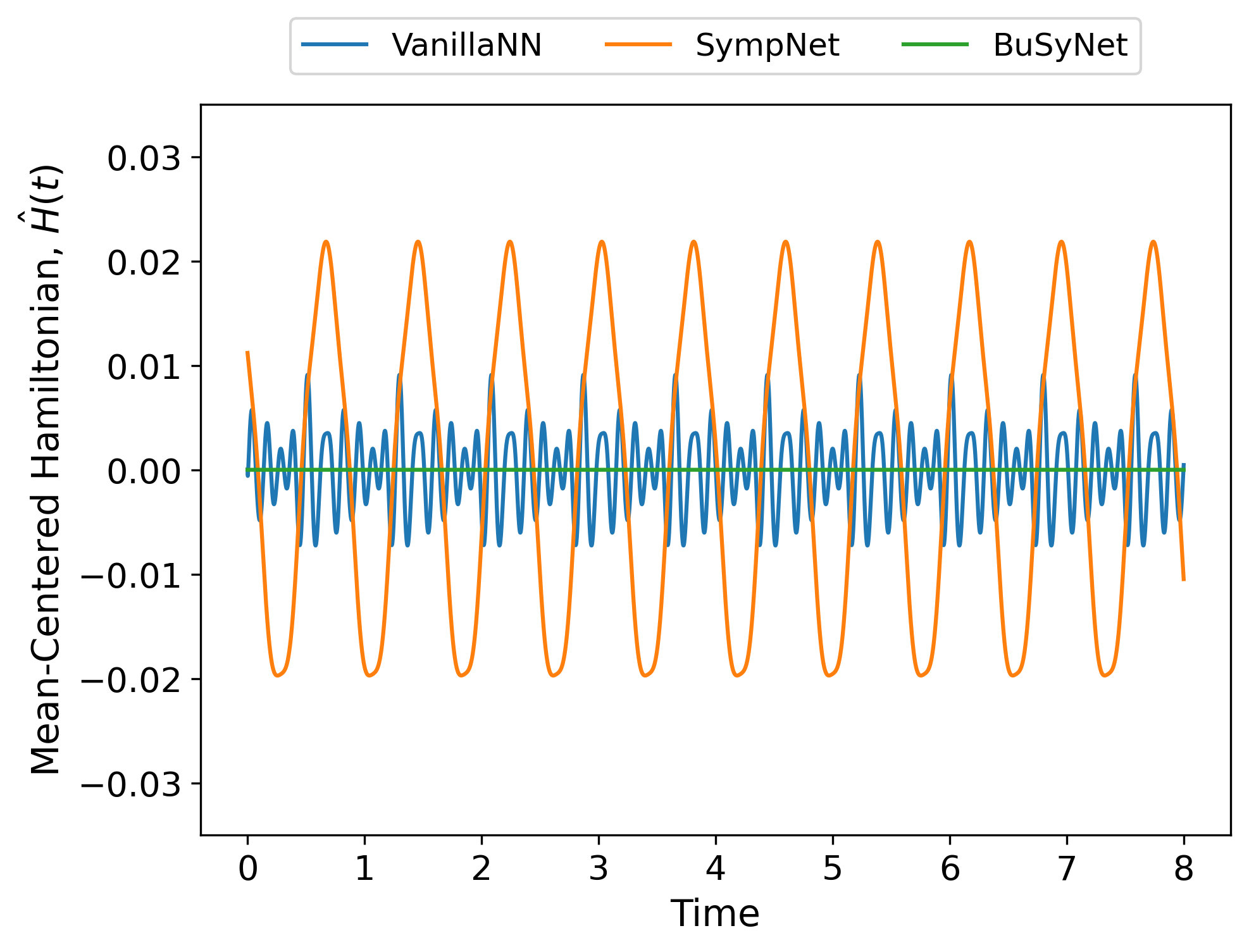}
        \caption{Mean-centered Hamiltonian for the 2D Kepler problem across the methods. HNN is also omitted.}
        \label{fig:KeplerHamiltonianPrediction}
    \end{subfigure}
  \end{minipage}
  \caption{Discovered Hamiltonian $\hat{H}$ versus time on the training and test sets for the BuSyNet method.}
  \label{fig:HamiltonianPredictions}
\end{figure}

\begin{figure}[htbp]
  \centering
  \begin{minipage}{\textwidth}
    \centering
    \begin{subfigure}[t]{0.46\textwidth}
        \centering
        \includegraphics[width=\textwidth]{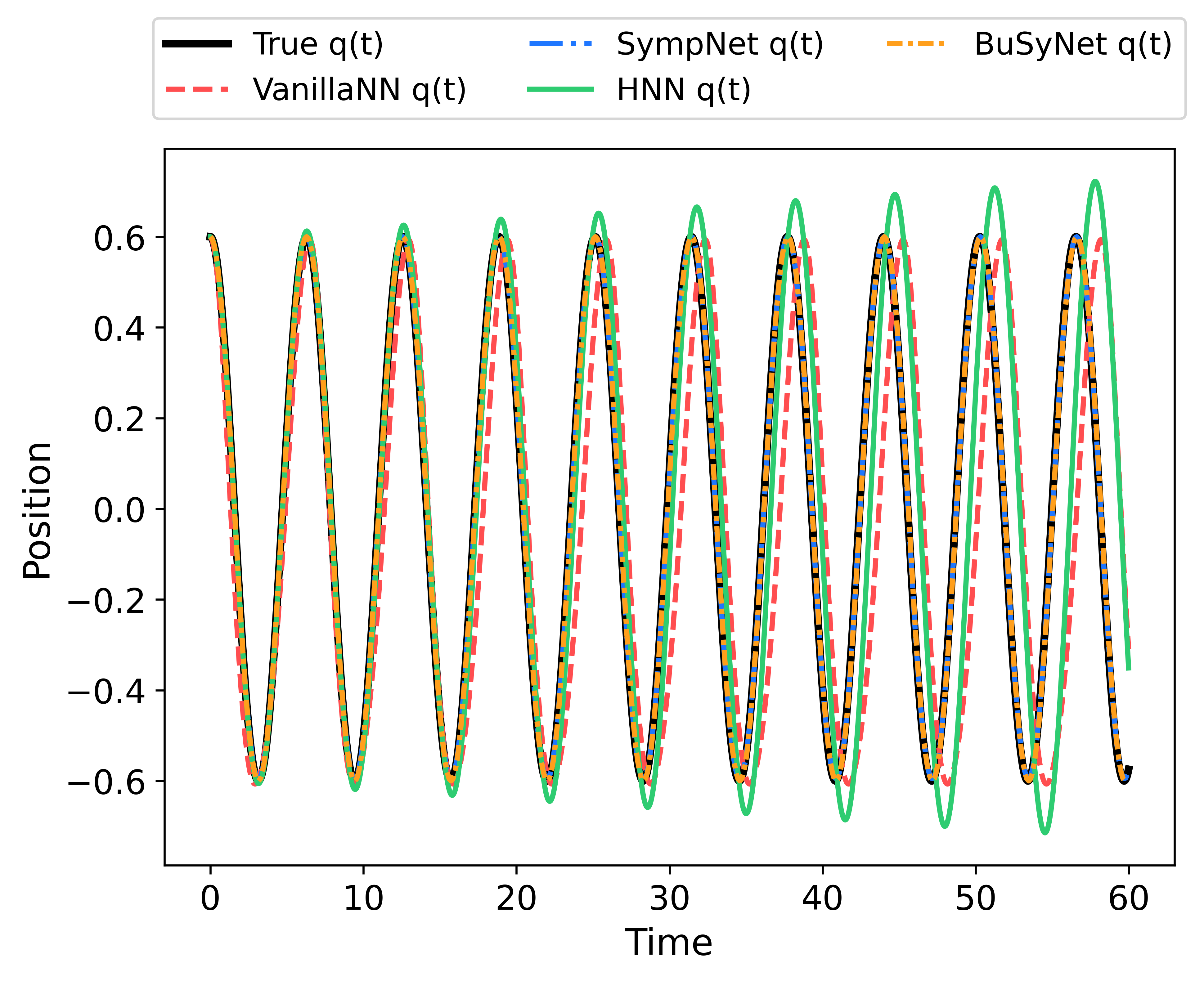}
        \caption{Comparison of methods for the simple harmonic oscillator}
    \end{subfigure}
    \hspace{0.05\textwidth}
    \begin{subfigure}[t]{0.46\textwidth}
        \centering
        \includegraphics[width=\textwidth]{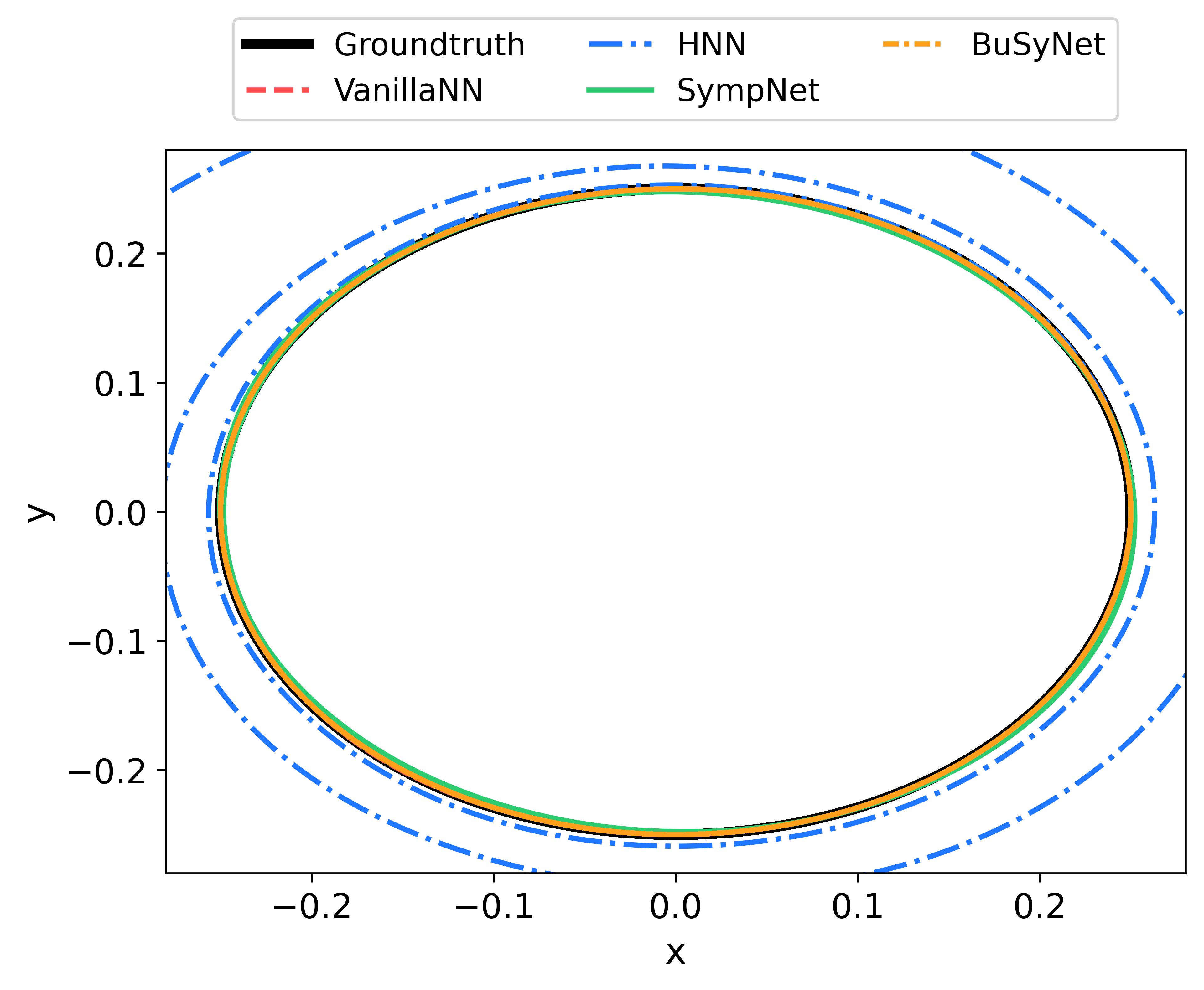}

        \caption{Comparison of methods for the Kepler problem in 2D, where $x=r \cos \theta, \ y=r \sin \theta$.}
        \label{fig:phase-comparison}
    \end{subfigure}
  \end{minipage}
  
  \caption{Graphs comparing the performance of BuSyNet with the other methods for the temporal evolution of the position $q(t)$ of the harmonic oscillator and the phase space (in Cartesian coordinates) for the 2D Kepler problem.}
  \label{fig:trajectoryPredictions_BuSyNet}
\end{figure}

Figure~\ref{fig:HamiltonianPredictions} plots the learned energy versus time for the simple harmonic oscillator; BuSyNet stays flat while NN, HNN, and SympNet gradually drift or oscillate around the true value. 
Energy drift over the entire time interval is negligible: for the oscillator, the variance of the Hamiltonian quantity is \(\operatorname{Var}\hat H=9.9\times10^{-12}\), outperforming other methods as shown in Figure \ref{fig:SHMHamiltonianPrediction}. 
As for the Kepler problem, energy variance is similarly very small at \(\operatorname{Var}\hat H=1.17 \times 10^{-11}\) for 2D and \(\operatorname{Var}\hat H=2.32 \times 10^{-11}\) for 3D. 
Long-term state predictions are shown in Fig.~\ref{fig:trajectoryPredictions_BuSyNet}; BuSyNet trajectories remain phase-locked with ground truth, whereas baselines accumulate more phase error.

Note that, since we are grounding the architecture with an explicit computation of the actions from trajectories, our architecture discovers the accurate numerical value of the Hamiltonian. Contrast this to other architectures like HNN, where the loss only enforces constraints on the gradient of the Hamiltonian, making those architectures only capable of identifying the value of the Hamiltonian up to a constant, showcasing another advantage of our approach. 

We obtain \(\hat{\omega}=1.0 = \omega_{\text{true}}\) for the harmonic oscillator.  For the Kepler system, the orbit corresponding to the chosen initial data is circular, and, accordingly, we recover the correct frequencies with $\hat{\omega}_{r}=0.0, \ \hat{\omega}_{\theta}=8.0$ in 2D and $\hat{\omega}_{r}=\hat{\omega}_{\theta}=0.0, \ \hat{\omega}_{\phi}=8.0$ in 3D, in exact agreement with the theoretical expectation. 

Figure~\ref{fig:trajectoryPredictions_BuSyNet} shows the evolution of the position for both the simple harmonic oscillator and the 2D Kepler problem (with $y=r \sin \theta$ versus $x=r \cos \theta$). BuSyNet reproduces trajectories that are indistinguishable from the ground truth compared to HNN and SympNet that diverge after a few periods of oscillation.

\begin{figure}[htbp]
  \centering
  \begin{minipage}{\textwidth}
    \centering
    \begin{subfigure}[t]{0.46\textwidth}
        \centering
        \includegraphics[width=\textwidth]{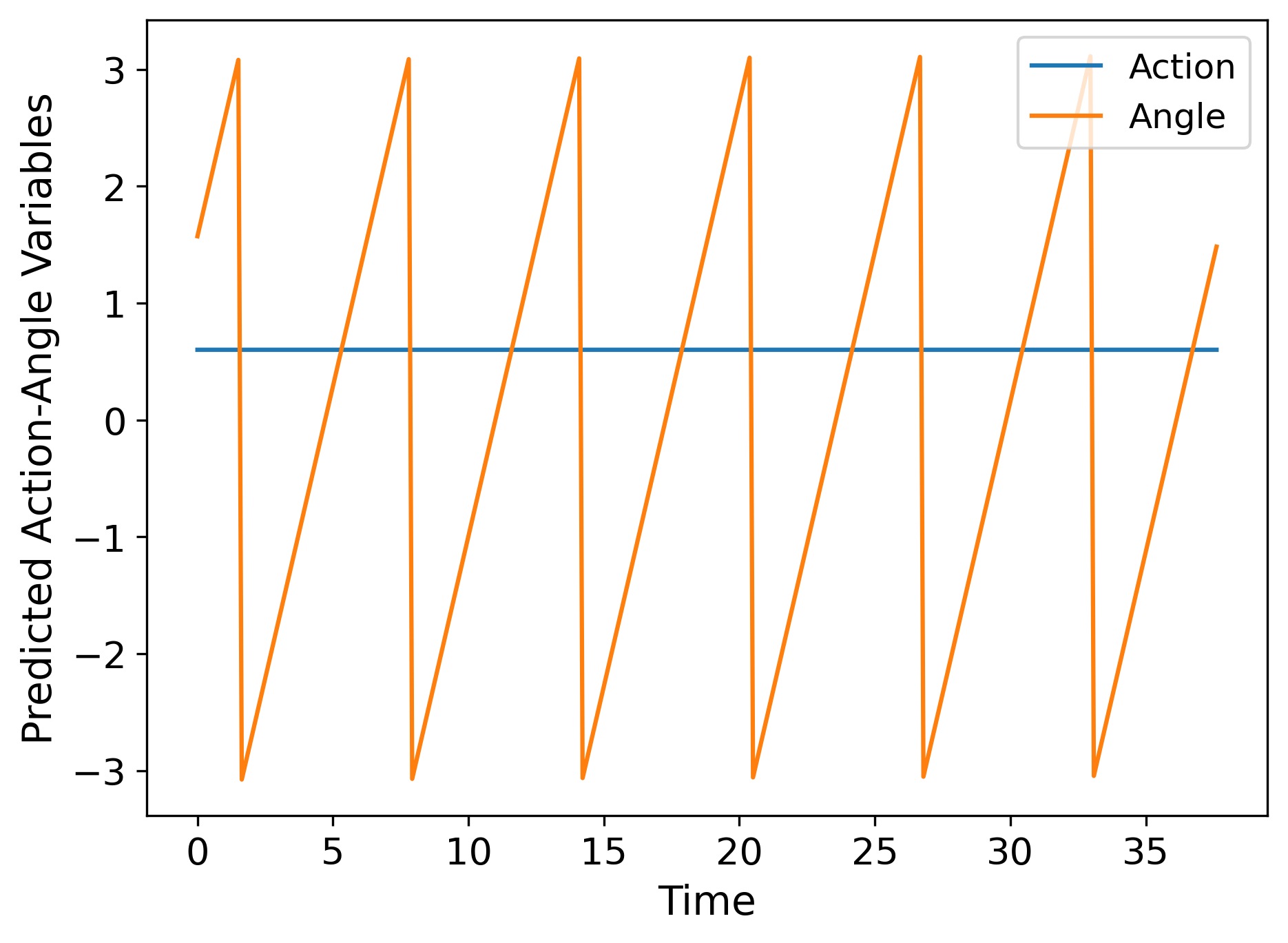}
        \caption{$I(t), \theta(t)$ for harmonic oscillator}
    \end{subfigure}
    \hspace{0.05\textwidth}
    \begin{subfigure}[t]{0.46\textwidth}
        \centering
        \includegraphics[width=\textwidth]{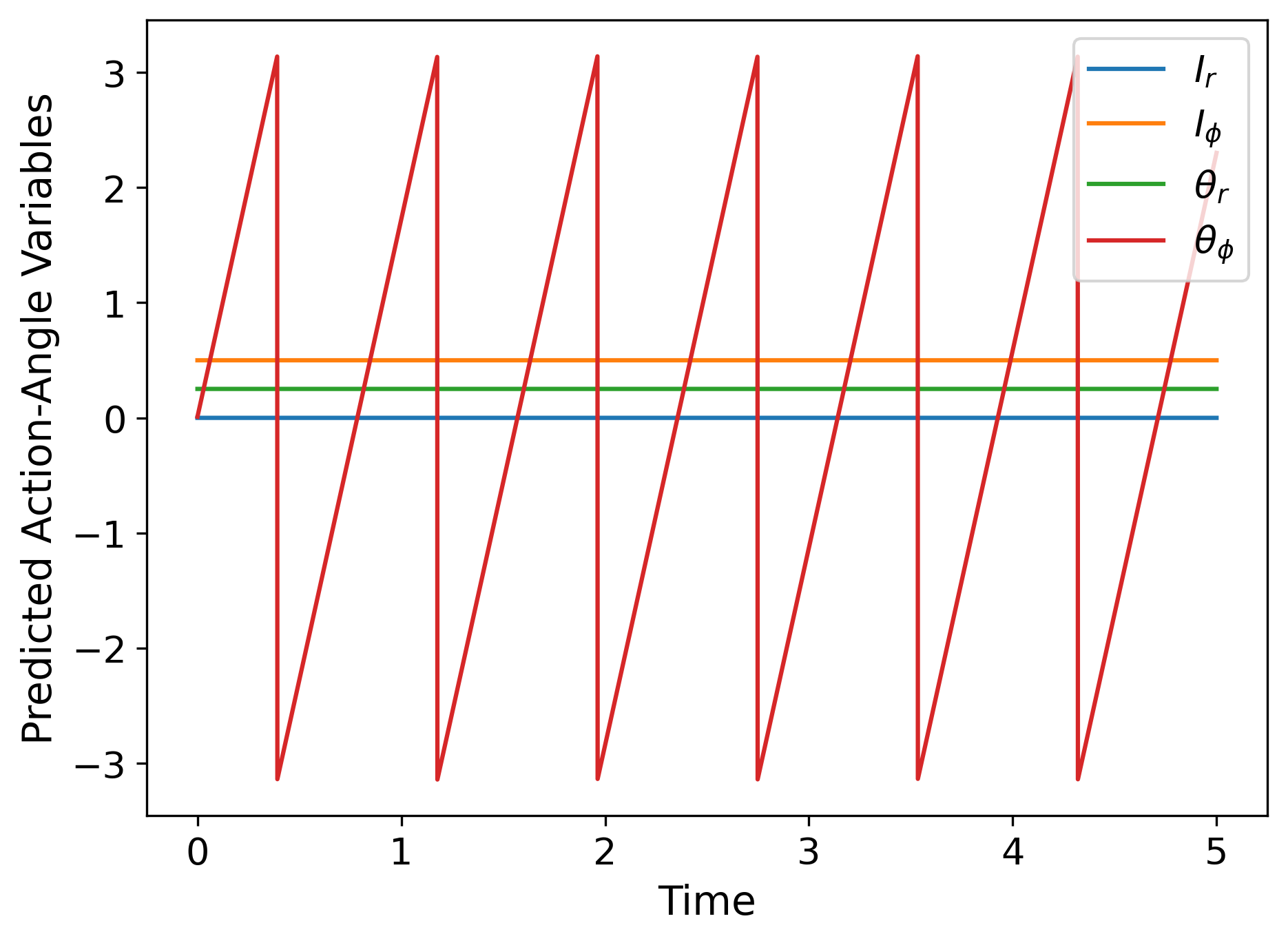}
        \caption{$\mathbf I(t), \boldsymbol \theta(t)$ for Kepler 2D problem}
    \end{subfigure}
  \end{minipage}
  \caption{The learned transformation to action-angle coordinates on the training and test sets for the BuSyNet method.}  \label{fig:ActionAnglePredictions}
\end{figure}

Figure~\ref{fig:ActionAnglePredictions} confirms that the learned transformation \(\mathcal S_\phi\) maps trajectories onto an invariant torus: actions $\mathbf I$ are constant, while angles increase linearly modulo \(2\pi\).  This validates the latent geometry that underpins BuSyNet's stability.

Tables \ref{tab:method_comparison} and \ref{tab:method_comparison_kepler} summarize mean-squared error (MSE) and energy variance results for the simple harmonic oscillator and the Kepler (2D and 3D) problems respectively. BuSyNet achieves the lowest error in every category and maintains five to six orders of magnitude better energy conservation than all baselines, signifying the key improvement brought about by learning an explicit form for the Hamiltonian and incorporating dimensional information in the network architecture.

\begin{table}[htbp]
\centering
\caption{Performance comparison on the 1D harmonic oscillator.}
\label{tab:method_comparison}
\begin{tabular}{lcccc}
\toprule
\textbf{Metric} 
& \textbf{Vanilla NN} 
& \textbf{HNN} 
& \textbf{SympNet} 
& \textbf{BuSyNet (Ours)} \\
\midrule

MSE on $[0,2T]$ 
& $1.73\times10^{-4}$ 
& $2.63\times10^{-4}$ 
& $1.74\times10^{-5}$ 
& $\mathbf{7.89\times10^{-6}}$ \\

MSE on $[2T,20T]$ 
& $3.12\times10^{-3}$ 
& $3.75\times10^{-1}$ 
& $3.95\times10^{-4}$ 
& $\mathbf{2.77\times10^{-4}}$ \\

$\operatorname{Var} \hat{H}(t)$ 
& $3.17\times10^{-3}$ 
& $3.10\times10^{-3}$ 
& $4.14\times10^{-6}$ 
& $\mathbf{9.88\times10^{-12}}$ \\

\bottomrule
\end{tabular}

\vspace{0.5em}
\small
The system parameters are $m=\SI{1}{\kilogram}$ and 
$k=\SI{1}{\kilogram\per\second\squared}$, yielding a period 
$T=2\pi\,\si{\second}$. 
Training data consists of $1000$ evenly spaced samples from 
$t=0$ to $t=\lceil 2\pi \rceil = \SI{7}{\second}$ with initial conditions 
$q(0)=\SI{0.6}{\meter}$ and 
$p(0)=\SI{0}{\kilogram\meter\per\second}$. 
Ground truth trajectories are generated using RK4 with step size $h=0.001$. 
For models that do not explicitly learn the Hamiltonian (Vanilla NN, SympNet), 
$\hat{H}(t)$ is computed using the analytical Hamiltonian for evaluation.
\end{table}

\begin{table}[htbp]
\centering
\caption{Performance comparison on Kepler systems.}
\label{tab:method_comparison_kepler}
\begin{tabular}{llcccc}
\toprule
\textbf{System} & \textbf{Metric} 
& \textbf{Vanilla NN} 
& \textbf{HNN} 
& \textbf{SympNet} 
& \textbf{BuSyNet (Ours)} \\
\midrule

\multirow{2}{*}{Kepler 2D}
& MSE over $[0,25T]$
& $5.20\times10^{-3}$ 
& $2.46\times10^{1}$ 
& $1.07\times10^{-4}$ 
& $\mathbf{5.25\times10^{-5}}$ \\

& $\operatorname{Var} \hat{H}(t)$ 
& $2.74\times10^{-5}$ 
& $6.49\times10^{2}$ 
& $9.92\times10^{-5}$ 
& $\mathbf{1.17\times10^{-11}}$ \\

\midrule

\multirow{2}{*}{Kepler 3D}
& MSE over $[0,25T]$
& $1.38\times10^{0}$ 
& $4.53\times10^{1}$ 
& $4.10\times10^{-5}$ 
& $\mathbf{3.51\times10^{-5}}$ \\

& $\operatorname{Var} \hat{H}(t)$ 
& $9.05\times10^{-6}$ 
& $8.30\times10^{3}$ 
& $3.09\times10^{-6}$ 
& $\mathbf{2.32\times10^{-11}}$ \\

\bottomrule
\end{tabular}

\vspace{0.5em}
\small
Similarly, for models that do not explicitly learn the Hamiltonian (Vanilla NN, SympNet), 
$\hat{H}(t)$ is computed using the analytical Hamiltonian for evaluation.
\end{table}

\section{Conclusion}

We have presented \textbf{BuSyNet}, a neural architecture that learns a canonical transformation to latent action-angle coordinates through a symplectic encoder and recovers a closed-form Hamiltonian with the correct energy units through a BuckiNet head. 
By embedding \emph{both} symplectic geometry and dimensional consistency in a differentiable pipeline, BuSyNet turns raw, time-varying trajectories into a single, time-invariant energy integral. 
On two canonical integrable systems, the harmonic oscillator and the Kepler problem, the network recovers symbolic Hamiltonians whose exponents match the analytic ground truth, while outperforming HNN, SympNet, and baseline neural networks baselines on long-horizon forecasting. The experiments demonstrate that respecting units narrows the hypothesis class and improves generalization, while the latent torus learned by the symplectic encoder eliminates drift and renders prediction almost trivial.

The present study still assumes full phase-space measurements; extending the method to position-only data, for instance by inferring momenta or by employing contact-Hamiltonian formalisms, remains an important direction. 
Likewise, BuSyNet currently targets integrable dynamics; adapting the architecture to weakly chaotic or fully non-integrable systems may require local action–angle charts or hybrid splittings. 
Finally, BuckiNet appears only at the output layer; propagating unit vectors through the entire depth of a network would yield fully dimension-aware models, a feature that could benefit turbulence, climate, and other multiscale applications.

Overall, this work shows that weaving dimensional analysis and symplectic structure into modern deep-learning pipelines is more than an aesthetic choice; it is a practical route to reliable, interpretable, and scientifically faithful machine-learning models.

\begin{acknowledgments}
The authors are grateful for helpful discussions with Profs. Mikhael Balabane and Jihad Touma. JB acknowledges support from the University Research Board.
\end{acknowledgments}


\newpage
\bibliographystyle{plainnat}  
\bibliography{bibliography.bib}


\end{document}